\documentstyle[epsf]{res}
\begin{document}

\title{%
Baryonic and Non-Baryonic Dark Matter}

\author{Bernard CARR\\
{\it Astronomy Unit, Queen Mary \& Westfield College,\\
Mile End Road, London E1 4NS, England, B.J.Carr@qmw.ac.uk}}

\maketitle

\section*{Abstract}

Cosmological nucleosynthesis calculations imply that there should be
both non-baryonic and baryonic dark matter. Recent data suggest
that some of the non-baryonic dark matter must be ``hot"
 (i.e. massive neutrinos) and there may also be evidence for ``cold"
dark matter (i.e. WIMPs). If the baryonic dark matter resides in galactic halos,
it is likely to be in the form of compact objects (i.e. MACHOs) 
and these would probably be the remnants of a first generation of 
pregalactic or protogalactic Population III stars. Many candidates
have been proposed - brown dwarfs, red dwarfs, 
white dwarfs or black holes - and at various times each
of these has been in vogue. We review the
many types of observations
which can be used to constrain or exclude both baryonic and non-baryonic 
dark matter candidates.

\section{Introduction}

Evidence for dark matter has been claimed in many different 
contexts. There may be {\em local\/} dark matter in the Galactic disc, dark matter in the {\em halos\/} of our own and other galaxies, dark matter associated with {\em clusters\/} of galaxies and finally - if one believes that the total
cosmological density has the critical value - smoothly 
distributed {\em background\/} dark matter. Since dark matter
probably takes as many different forms as visible matter, it would be simplistic
to assume that all these dark matter problems have a single solution. The local
dark matter is almost certainly unrelated to the other ones and, while the halo and
cluster dark matter would be naturally connected in many evolutionary scenarios, there is a growing tendency to regard
the unclustered background dark matter as different from the clustered component. 
This is because the latest supernovae measurements indicate that the
cosmological expansion is accelerating, which means that the total density must be
dominated by some form of ``exotic" energy with negative pressure (possibly a cosmological constant). 

The combination of the supernovae and microwave background observations suggest
that the density parameters associated with the exotic (unclustered) component and the ordinary (positive pressure) {\em matter\/} component are $\Omega_X=0.8\pm0.2$ and $\Omega_M=0.4\pm0.1$ respectively (Turner 1999). This is compatible with the total density parameter being $1$, as expected in the inflationary scenario, but does not
definitely require it. This paper will focus exclusively on the matter component of
the cosmological density, with particular emphasis on the question of how much of this is baryonic and non-baryonic. We will also be interested in the relative contributions of
hot and cold non-baryonic matter. As emphasized by Turner (1999), each of these components seem to be required and it is remarkable that all their densities are within one or two orders of magnitude of each other. Why this should be remains a mystery but we will not explore this further here. 

\section{Evidence for Baryonic and Non-Baryonic Dark Matter}

The main argument for both baryonic and
non-baryonic dark matter comes from Big Bang nucleosynthesis calculations. 
This is because the success of the standard picture in explaining the
primordial light element abundances only applies if the baryon 
density parameter lies in the range (Copi et al. 1995)
\begin{equation}
      0.007h^{-2} < \Omega_B < 0.022h^{-2}
\end{equation}
where $h$ is the Hubble parameter in units of
$100$~km~s$^{-1}$~Mpc$^{-1}$. 
A few years ago the anomalously high deuterium abundance measured in some intergalactic Lyman-$\alpha$ clouds suggested that the
nucleosynthesis value for $\Omega_B$ could be lower than indicated by eqn (1) (Carswell et al. 1994, Songaila et al. 1994, Rugers \& Hogan 1996, Webb at al. 1997). However, the evidence for this has always been strongly disputed (Tytler et al. 1996) and recent studies of the
quasars Q1937 and Q1009 suggest that the deuterium abundance is $3.3 \times 10^{-5}$ (Burles \& Tytler 1998). This corresponds to the range
\begin{equation}
      0.018h^{-2} < \Omega_B < 0.020h^{-2},
\end{equation}
which is much tighter than indicated by eqn (1) (Burles et al. 1999). 

Both eqns (1) and (2) suggest that the upper limit on $\Omega_B$ is well below 1, which implies that no
baryonic candidate could provide the matter density required by large-scale
structure observations. This conclusion also applies if one invokes
inhomogeneous nucleosynthesis since one requires $\Omega_B<0.09h^{-2}$ even
in this case (Mathews et al. 1993). On the other hand, the upper limit
on $\Omega_B$ allowed by eqns (1) and (2) almost certainly exceeds the density of 
visible baryons $\Omega_V$. A careful inventory by Persic \& Salucci (1992)
shows that the density in galaxies and cluster gas is 
\begin{equation}
\Omega_V \approx (2.2 + 0.6h^{-1.5}) \times 10^{-3} \approx 0.003
\end{equation}
where the last approximation applies
for reasonable values of $h$. A recent review by 
Hogan (1999) has updated this estimate of $\Omega_V$: he replaces the factors in brackets in eqn (3) by 2.6 for spheroid stars, 0.86 for disc stars, 0.33 for neutral atomic gas, 0.30 for
molecular gas and 2.6 for the ionized gas in clusters (assuming 
$H_o =70$~km~s$^{-1}$~Mpc$^{-1}$). Although this is still well below the nucelosynthesis bound, this may not account for all components. Some of the baryons may be in an intergalactic medium; for example,  Cen \& Ostriker (1999)
find that - in the context of the ``Cold Dark Matter" scenario - half the mass of baryons should be in warm ($10^5-10^7$K) intergalactic gas at the present epoch. 
Another possibility is that the missing baryons
could be in gas in groups of clusters. This has been emphasized by Fukugita et al. (1998), who argue 
that plasma in groups (which would be too cool to have appreciable X-ray emission)
could provide all of the cosmological nucleosynthesis shortfall. Indeed the review by Hogan (1999)
suggests that the ionized gas in groups could have a density parameter as high as $0.014$. However, it must be stressed that this estimate is based on an extrapolation of observations of rich clusters and there is no direct evidence for this.

Which of the dark matter problems identified in Section 1 would be baryonic or non-baryonic? We have already seen that the dark matter in clusters would need to be 
mainly non-baryonic. On the other hand, the dark matter in galactic discs could certainly be baryonic: even if all discs have the 60\% dark component
envisaged for the Galaxy by Bahcall et al. (1992), this only 
corresponds to $\Omega_d \approx
0.001$, well below the nucleosynthesis bound. Indeed the disc dark matter would 
{\it need} to be baryonic in order to dissipate enough to form a disc. However, the Bahcall et al. claim is strongly disputed by Kuijken \& Gilmore (1989) and Flynn \& Fuchs (1995), In fact, recent Hipparcos observations suggest that the dark disc fraction is below 10\% (Cr\'ez\'e et al. 1998).

The more interesting question is whether the 
halo dark matter could be baryonic. If the Milky Way is typical, the density associated with halos would be 
\begin{equation}
\Omega_h \approx 0.03h^{-1}(R_h/100\mbox{kpc}),
\end{equation}
where $R_h$ is the (uncertain) halo radius, so the upper limit in eqn (1) 
implies that {\em all\/} the dark matter in halos could be baryonic 
providing
$R_h<70h^{-1}$kpc. This is marginally possible for our galaxy (Fich \& 
Tremaine 1991), in which case the dark baryons could be contained in the remnants of a first generation of ``Population III" stars (Carr 1994). This corresponds to the ``Massive Compact Halo Object" or ``MACHO" scenario and has attracted considerable interest as a result of the LMC microlensing observations. Even if halos are larger than $70h^{-1}$kpc, and studies of the kinematics of other spirals by Zaritsky et al. (1993) suggest that they could be as large as $200h^{-1}$kpc, a considerable fraction of their mass could still be in stellar remnants.

On theoretial grounds one would expect the halo dark matter to be a {\it mixture} of WIMPs and MACHOs. For since the cluster dark matter must be predominantly cold, one would expect at least some of it to clump into galactic
halos. The relative densities of the two components would depend sensitively on
the formation epoch of the Population III stars. If they formed pregalactically, one would expect
the halo ratio to reflect the cosmological background ratio (viz. $\Omega_B/\Omega_M \approx 0.1$). However, if they formed protogalactically, the ratio could be larger
since the baryons could have dissipated and become more centrally concentrated 
than the WIMPs.
 In order to distinguish between the pregalactic and protogalactic scenarios, it is important to gain independence evidence
about the formation epoch of the putative MACHOs. At moderate redshifts one can obtain a lower limit to the baryon density by studying Lyman-$\alpha$ clouds. The simulations of Weinberg et al. (1997) suggest that the
density parameter of the clouds must be at least $0.017h^{-2}$ at a redshift of 2 and this is already
close to the upper limit given by eqn (1). By today some of these baryons might have been transformed into a hot intergalactic medium or stars but this suggests that there was little room for any dark baryons before $z=2$. On the other hand, we will see
in Section 4 that background light constraints require that any massive Population III stars must have formed much earlier than this.

\section{Non-Baryonic Candidates}

Not only is the presence of non-baryonic dark matter
required by cosmological nucleosynthesis arguments, but it also {\it expected} in the Big Bang scenario. This is because any particles which exist in nature must have existed profusely in the early Universe once the temperature exceeded their rest mass $m_X$. Some of
these particles should therefore have survived as ``relics" once  the temperature fell below the value $T_F$ at which their interactions froze-out. If the particles were relativistic at freeze-out ($T_F>m_X$), their current number density should be comparable to that of the microwave background photons, so one expects a relic density parameter $\Omega_X \approx (m_X/100eV)$. On the other hand, 
if they were non-relativistic at freeze-out ($T_F<m_X$), the current density declines with the particle mass ($\Omega_X \sim m_X^{-1.9}$) due to Boltzmann suppression. This gives two mass windows in which $\Omega_X$ can be interesting:
$M_X\approx 10-100$ eV and $M_X\approx 10-30$ GeV. 
Since the current velocity dispersion of the particles should be 
$<v^2>^{1/2}\approx T/m_X$, which decreases with mass,
particles in the smaller and larger ranges are termed ``hot" and ``cold" respectively. 
Particles can clump into bound objects of radius $R$ and velocity dispersion $\sigma$ only for 
\begin{equation}
m_X > 30(R/10kpc)^{-1/2}(\sigma/100km/s)^{-1/4}eV
\end{equation}
 and this gives a lower limit
of 4 eV for clusters, 20 eV for galactic halos and 300 eV for dwarf galaxies.

Large-scale structure observations also require that the Universe be dominated by non-baryonic matter. This is because, in a purely baryonic universe, the density fluctuations required at decoupling to generate this structure would induce larger microwave background anisotropies than observed. However, this argument does not specify the nature of the non-baryonic dark matter precisely. 
The original ``hot dark matter" (HDM) scenario, proposed in the early 1980s, assumed that it was in the form of neutrinos but that would be incompatible with the presence of dark matter in dwarf galaxies. This was superseded in the mid-1980s by the ``cold dark matter" (CDM) scenario and this remained popular for almost a decade when it was found that
it did not  fit the latest  large-scale structure data or give the observed gas fraction in clusters. In the mid-1990s models invoking a mixture of hot and cold dark matter (CHDM) became topical but these still had problems explaining the observed abundances of clusters, bright galaxies and Lyman-$\alpha$ clouds. 
Currently we have seen that a Universe containing both a cosmological constant and cold dark matter ($\Lambda CDM$) is in vogue and some people have considered a hot component ($\Lambda CHDM$) as well.

Eventually the combination of large-scale structure data and microwave background 
anisotropy measurements could determine the relative contributions of all the different types of dark 
matter very precisely (Gawiser \& Silk 1998). However, as mentioned in the 
Introduction, it already seems clear that each of the components is needed and that their densities are all comparable.
Of course, these arguments provide only indirect evidence for non-baryonic
dark matter. More definitive evidence can come only from attempts to detect the hot or cold particles directly. There have been several recent interesting developments in this context, as we now discuss.
\vspace{10pt}

{\em Massive Neutrinos}. There are now three experiments which indicate a non-zero neutrino mass and hence a non-zero cosmological neutrino density [see Caldwell (1999) for a recent review].
(1) Studies of the ratio of muon and electron neutrinos produced
by the decay of pions in the atmosphere suggest that there are oscillations between  muon and tau neutrinos, corresponding to a mass difference
$\Delta m_{\tau \mu}^2 \approx 2 \times 10^{-3}eV^2$.
(2) The deficit of electron-neutrinos coming from the Sun implies that these neutrinos must be 
oscillating into another neutrino species (labelled ``X") with a mass difference, if attributed to the MSW effect, given by 
$\Delta m_{eX}^2 \approx 10^{-5}eV^2$.
(3) LSND claims to have detected oscillations between muon and electron neutrinos in neutrino beams coming from accelerator experiments, corresponding to a mass difference
$ 10 eV^2 > \Delta m_{\mu e}^2> 0.2 eV^2$. This is a controversial result but, if true, it implies $m(\nu_\mu)>0.5eV$ 
and hence $\Omega_{\nu}>0.005$, which is already comparable to the visible density given by eqn (3). Since the first experiment implies  
$\Delta m_{\tau e}^2 >> 10^{-5} eV^2$, the particle X required to explain the 
solar neutrino deficit cannot be the tau nuetrino and so would most naturally be identified with the muon neutrino. In this case,
\begin{equation}
m(\nu_\tau) \approx \Delta m_{\tau \mu}\approx 0.05eV, \;\;\; m(\nu_\mu)\approx \Delta m_{\mu e}\approx 0.005eV, \;\;\; m(\nu_e) << 0.005eV.
\end{equation}
This is consistent with the ``see-saw" mechanism and suggests that the total neutrino density is very low. Another possibility, and the only one
compatible with the LSND result,  is that  $m(\nu_\tau) \approx m(\nu_\mu)$, so that one has mass degeneracy and $m(\nu_\tau) >> \Delta m_{\tau \mu}$. In this case $\Omega_{\nu}$ could be large and is limited only by the large-scale structure observations (i.e. $\Omega_{\nu}<0.2$).
However, the solar neutrino problem could not then be atributed to electron neutrinos oscillating into muon neutrinos since $ \Delta m_{\mu e}$ would be too large. One would therefore need to identify X with a new ``sterile" neutrino, whose mass is very close to that of the electron neutrino, so that  $\Delta m_{\mu e} >> \Delta m_{X e}$.
It has been claimed that the sterile neutrino may be useful for astrophysical purposes 
(eg. in explaining the abundance of r-process elements) but this remains controversial.
\vspace{10pt}

{\em WIMPs}.  Supersymmetry theory requires the existence of a large number 
extra ``sparticles" with a mass in the range $0.1-1$ TeV. [See Roszkowski
(1999) and Ellis (2000) and for recent reviews.] The Lightest Supersymmetric Particle (LSP) should be stable and observations require that it has neither electric nor
strong interactions. Three candidates have been suggested: the (spin 0) sneutrino, the (spin 3/2) gravitino, and the (spin 1/2) neutralino. However, the LEP results imply that the sneutrino mass must exceed $1$ TeV, which gives a negligible density, while the gravitino would have a mass of order $1$ keV, corresponding to ``warm" dark matter. The latter would be inconsistent with large-scale structure observations, although
another scenario has recently been proposed in which 100 GeV gravitinos are produced in the context of leptogenesis.
The neutralino is therefore the favoured WIMP candidate and this would need to be some mixture of the photino, Higgsino and zino. LEP implies that the mass must be in the range $30-600$ GeV, corresponding to $\Omega_Xh^2 \sim 0.1$. One would expect there to be a continual flux of halo neutralinos passing through the Earth and efforts to detect these has been underway for more than a decade.   
Recently the DAMA group claims to have made a positive detection, associated with the
annual modulation of the flux as the Earth goes around the Sun. They find a mass in the range $45-76$ GeV but it should be stressed that this depends sensitively on the halo model.
\vspace{10pt}

{\em Axions}. The axion is a particle associated with the breaking of Peccei-Quinn symmetry
at a temperature $f_a \approx 10^{12}$GeV and it aquires a mass $m_a$ due to 
instanton effects at a
temperature $T\approx \Lambda_{QCD}$. In the original thermal production
scenario, the associated parameters are
\begin{equation}
\Omega_a \approx \theta^2(m_a/10^{-6}eV)^{-1.2},\;\;\ m_a \approx  \Lambda_{QCD}^2/f_a \approx 10^{-5}(f_a/10^{12})^{-1}eV
\end{equation}
where $\theta$ is the ``misalignment" angle (Preskill et al. 1983). Several astrophysical constraints require
$10^{-6}eV<m_a<10^{-2}eV$ and microwave cavity searches focus on the mass range $10^{-6}-10^{-5}eV$. Later it was realized that a network of axion strings could form at the Peccei-Quinn epoch (Davis \& Shellard 1989) and axion domain walls
could form at the QCD epoch (Shellard 1990). The decay and interaction of these strings and walls could produce an axion density in the range 
$\Omega_a\approx (100-1000)(m_a/10^{-6}eV)^{-1.2}$, which is considerably
larger than  indicated by eqn (7). Various refinements of these
scenarios arise in the inflationary scenarios, depending on whether inflation occurs at a higher or lower temperature than $f_a$ (Shellard \& Battye 1998). The combination of the Supersymmetry 
and Peccei-Quinn symmetry also implies the existence of an axino. In the simplest scenario this has a mass of order a keV and therefore acts like warm dark matter. However, in a more recent scenario the photino decays into an axino and a
photon and this predicts a much larger axino density (Covi et al. 1999). 
\vspace{10pt}

{\em Superheavy Relics}. The existence of WIMPs is an inevitable prediction of the ``bottom-up" approach of supersymmetry theory. However, one can also adopt various
``top-down" approaches (eg. heterotic string theory and minimal supergravity) and these predict the existence of a plethora
of particles with much larger mass. For example, Kaluza-Klein theory predicts many helicity states with mass near the Planck mass (Ellis et al. 1990) and the non-thermal decay of inflatons would produce a large density of GUT-mass particles (Kolb 2000).  Such particles are
sometimes described as ``wimpzillas" and their decay might 
provide a possible explanation of the Ultra-High Energy (UHE) cosmic rays with $E>10^{19}$GeV. The origin of such cosmic rays is currently unknown but they must originate within 30-50 Mpc (because of the GZK effect) and they do not seem to be associated with individual galaxies or gamma-ray bursts. Another possible explanation would be the decay of topological defects. In any case, observations of UHE cosmic rays already place interesting constraints on such scenarios (Ziaeepour 1999). 
\vspace{10pt}

{\em Indirect Detections}. There are various ways in which
dark matter particles might be detected indirectly. For example, the annihilations of WIMPs in
the Galactic halo would produce a flux of positron and antiproton cosmic rays
(Bergstrom 1999). Balloon experiments have detected an excess of such particles
but it is difficult to identify them specifically with WIMPs. They might also be 
secondary particles produced by the interaction of primary cosmic rays with interstellar
protons and there is much uncertainty associated with the propagation of
cosmic rays through the Galaxy and their modulation by the solar wind. 
WIMP-annihilations would also induce a a Galactic $\gamma$-ray background, with a characteristic 
enhancement of a $\gamma$-line near
the Galactic centre in the CDM scenario. Such a background may have already been detected with the EGRET satellite (Chary \& Wright 1998).

\section{Baryonic Candidates}

Over the last few decades, a number of observational claims have been made from time
to time which seemed to indicate the existence of MACHOs. Many of these observations turned out to be spurious but they served the useful purpose of stimulating theorists to study cosmological
scenarios involving baryonic dark matter and to try to predict their various signatures. Currently one cannot state definitely that MACHOs exist but one can place 
important constraints on the possible candidates. In this section we will discuss
these candidates in turn, focussing particularly  on brown dwarfs, red dwarfs, white dwarfs and black holes.
\vspace{10pt}

{\em Background Radiation from VMOs}. Stars in some range above $8~M_\odot$ would leave neutron star or black hole remnants but neither of these would be plausible candidates for either the disc or halo dark matter because their precursors would have unacceptable nucleosynthetic yields. 
Stars larger than $200~M_\odot$ are termed ``Very Massive Objects" or VMOs and might 
collapse to black holes without these nucleosynthetic consequences (Carr et al. 1984). However, during their main-sequence phase, such VMOs would be expected to generate of a lot of background light. By today this should have been shifted into the infrared or submillimetre band, as a result of either redshift effects or dust reprocessing, so one would expect a sizeable infrared/submillimetre cosmic background (Bond et al. 1991). Over the last few decades there have been 
several reported detections of such a background but these have usually turned out to be false alarms. COBE does now seem to have detected a
genuine infrared background (Fixsen et al. 1998) but this can probably be attributed to 
ordinary Population I and II stars. In any case, the current constraints on such a background strongly limit the density of any VMOs unless they form at a very high redshift.  For this reason massive Population III stars would need to be pregalactic rather than protogalactic. 
\vspace{10pt}

{\em Dynamical Effects of SMOs}. Stars larger than $10^5~M_\odot$ - termed ``Supermassive Objects" or SMOs - would collapse directly to black holes without any nucleosynthetic or background light 
production. However, supermassive black holes would still have noticeable
dynamical effects and these have been investigated  by
many authors. The constraints on black holes in our own disc - due to the disruption of open clusters - and in our own halo - due to the heating of disc stars, the disruption of globular clusters and dynamical friction effects - have been discussed in detail by Carr \& Sakellariadou (1998). Although it has been claimed that there is positive evidence for some of these effects, such as disc heating (Lacey \& Ostriker 1985), the interpretation of this evidence is not clear-cut. 
It is therefore more natural to regard these dynamical effects  as merely imposing an upper limit on the density of black holes or indeed any other type of compact object.
It should be
stressed that many of these limits would also apply if the black holes were replaced by ``dark clusters"
of smaller objects, a scenario which has been explored by many authors (Carr \& Lacey 1987, Ashman 1990, Kerins \& Carr 1994, De Paolis et al. 1995, Moore \& Silk 1995).
\vspace{10pt}

{\em Halo Light from Red Dwarfs}. A particularly interesting recent development was the discovery of red light around NGC 5907 by Sackett et al. (1994), apparently emanating from low mass stars with a density profile like that of the halo. This detection was confirmed in V and I by Lequeux et al. (1996) and in J and K by James \& Casali (1996). However, the suggestion that
the stars might be of primordial origin (with low metallicity) was contradicted by the results of Rudy et al. (1997), who found that the color was indicative of low mass stars with solar metallicity.
Furthermore, it must be stressed that the red light light has only been observed within a few kpc and no NIR emission is detected at 10-30 kpc (Yost et al. 1999). Both these points go against the suggestion that the red light is associated with MACHOs.
Recently it has been suggested that the red light seen
in NGC 5907 is more likely to derive from a disrupted dwarf galaxy, the stars of which
would naturally follow the dark matter profile (Lequeux et al. 1998), or to be a ring left over from a disrupted dwarf spheroidal galaxy (Zheng et al. 1999). In any case, NGC 5907 does not seem to be typical since ISO observations of four other edge-on bulgeless spiral galaxies give no evidence for red halos (Gilmore \& Unavane 1998).
\vspace{10pt}

{\em Microlensing of Macrolensed Quasars by LMOs}. If a galaxy is 
suitably positioned to 
image-double a quasar, then there is also a high probability that 
an individual halo object will traverse the line of sight of one of the
images and this will give intensity fluctuations in one
but not both images (Gott 1981). The effect would be observable for objects 
bigger than $10^{-4}M_\odot$ but the timescale of the fluctuations
$\sim 40(M/M_\odot)^{1/2}$y would make them detectable only for  
$M<0.1~M_\odot$, so this method works only for ``Low Mass Objects" or LMOs. There is already evidence of this effect for the 
quasar 2237+0305 (Irwin et al. 1989), the observed
timescale for the variation in the luminosity of one of the images
possibly indicating a mass below $0.1~M_\odot$ (Webster et al. 1991). However, because the optical depth is high, the mass estimate is very uncertain and
a more recent analysis suggests that it could be in the range
$0.1 - 10~M_\odot$ (Lewis et al. 1998), in which case the lens could be an
ordinary star. The absence of this effect in the quasar 0957+561 has also been used to exclude MACHOs with mass in the range $10^{-7} - 10^{-3}M_\odot$ from making up all of the halo of the intervening galaxy, although the precise limit has some dependence on the quasar size (Schmidt \& Wambsganss 1998).
\vspace{10pt}

{\em Microlensing of Quasars by Jupiters}. More dramatic 
but rather controversial evidence for the microlensing of quasars 
comes from Hawkins (1993, 1996, 1999), who has been monitoring 300 quasars in 
the redshift range $1-3$ for nearly 20 years using a wide-field 
Schmidt camera. He finds quasi-sinusoidal variations with an amplitude of 
0.5 magnitudes on a timescale 5~y and attributes this to lenses with 
mass $\sim 10^{-3}M_\odot$. The crucial point is that the timescale 
decreases with
increasing redshift, which is the opposite to what one would 
expect for intrinsic variations, although this has been disputed (Alexander 1995, 
Baganoff \& Malkan 1995). The timescale also increases with the luminosity of the quasar and he explains this by noting that the
variability timescale should scale with the size of the accretion 
disc (which should itself correlate with luminosity). A rather worrying
feature of Hawkins' claim is that he requires the density of the
lenses to be close to critical (in order that the sources are
transited continuously), so he has to invoke primordial black 
holes which form at the quark-hadron phase transition 
(Crawford \& Schramm 1982). 
However, this requires fine-tuning since the fraction of the
Universe going into black holes at this transition must only be about $10^{-9}$. Walker (1999) has proposed that Hawkins' lenses might also be jupiter-mass gas clouds.  
\vspace{10pt}

{\em Microlensing of Stars in the LMC}. 
Attempts to detect microlensing by objects in our own halo by 
looking for intensity variations in stars in the Magellanic Clouds and the
Galactic Bulge have now been underway for a decade (Paczynski 1996). This method
is sensitive to lens masses in the range $10^{-7}-10^2M_\odot$ but the probability of an
individual star being lensed is only $\tau \sim 10^{-6}$, so one 
has to
look at many stars for a long time (Paczynski 1986). The duration and likely 
event rate are
\begin{equation}
P\sim 0.2(M/M_\odot)^{1/2}y,\quad \Gamma \sim N \tau P^{-1} \sim (M/M_\odot)^{-1/2}y^{-1}
\end{equation}
where $N\sim 10^6$ is the number of stars. The MACHO 
group currently has 13-17 LMC events and the durations span the range 
$34-230$ days (Alcock et al. 
2000). For a standard halo model, the data suggest an average lens mass of around $0.5~M_\odot$ and a halo fraction of 0.2, with the 95\% confidence ranges being $0.15-0.9~M_\odot$ and $0.08-0.5$. The mass is comparable with the earlier estimates but the
fraction is somewhat smaller (Alcock et al. 1997). This
might appear to indicate that the MACHOs are white
dwarfs but one could also invoke a less conventional candidate. For
example, primordial black holes forming at the
quark-hadron phase transition might have the required mass 
(Jedamzik 1997) and the microlensing implications of this scenario are
discussed by Green (2000). Perhaps the
most important result of the LMC searches is that they {\em eliminate} many candidates. Indeed the combination of the MACHO and EROS results already excludes objects in the mass range $5\times10^{-7}-0.002~M_\odot$ from having more than 0.2 of the halo density (Alfonso et al. 1997).
\vspace{10pt}

{\em White Dwarfs}. A few years ago white dwarfs (WDs) were regarded as rather implausible dark matter candidates. One required a very contrived IMF, lying between $2~M_\odot$ and $8~M_\odot$, 
in order to avoid excessive production of light or metals (Ryu et al. 1990); the fraction
of WD precursors in binaries would be expected to produce too many type 1A supernovae (Smecker \& Wyse 1991); and the halo fraction was constrained to be less than 10\% in order to avoid the luminous precursors contradicting the upper limits from galaxy counts (Charlot \& Silk 1995).  The observed WD luminosity function
also placed a severe lower limit on the age of any WDs in our own halo (Tamanaha et al. 1990). More recent constraints - from CNO production (Gibson \& Mould 1997), helium and deuterium production (Field et al. 2000) and extragalactic
background light limits (Madau \& Pozzetti 1999) - have strengthened these limits (Graff et al. 1999).
At least some of these limits must be reconsidered in view of recent claims by
Hansen (1998) that metal-poor old WDs with hydrogen envelopes could be much bluer and brighter than previously supposed, essentially because the light emerges from deeper in the atmosphere. 
This suggestion has been supported by HST observations of Ibata et al. (1999), who claim to have detected five candidates of this kind. The objects are blue and isolated
and show high proper motion. They infer that they are $0.5~M_\odot$ hydrogen-atmosphere WDs with an age of around $12$ Gyr. Three such objects have now been 
identified spectroscopically (Hodgkin et al. 2000, Ibata et al. 2000), so this possibility
must be taken very seriously. However, this does not circumvent the nucleosynthetic  arguments against WDs. 
\vspace{10pt}

{\it Brown Dwarfs}. Objects in the range $0.001-0.08~M_\odot$ would never burn hydrogen and are termed ``brown dwarfs" (BDs). They represent a balance between gravity and degeneracy pressure.
It has been argued that objects below the hydrogen-burning 
limit  may form efficiently in pregalactic or protogalactic cooling flows (Ashman \& Carr 1990, Thomas \& Fabian 1990) but the direct evidence for such objects
remains weak. While some BDs have been found as
companions to ordinary stars, these can only have a tiny cosmological density and 
it is much harder to find isolated field BDs. The best argument therefore comes from 
extrapolating the initial mass function (IMF) of hydrogen-burning stars to lower masses than can be observed directly. The IMF for Population I stars ($dN/dm \sim m^{-\alpha}$ with $\alpha<1.8$) suggests that only 1\% of the disc could be in BDs (Kroupa et al. 1995). However, one might wonder whether $\alpha$ could be larger, increasing the BD fraction, for zero-metallicity stars. Although there are theoretical reasons for entertaining this possibility, earlier observational claims that low metallicity  objects have a steeper IMF than usual are now discredited. Indeed observations of 
Galactic and LMC globular clusters (Elson et al. 1999) and dwarf spheroidal field stars
(Feltzing et al. 1999) suggest that the IMF is {\it universal} with $\alpha <1.5$ at low masses (Gilmore 1999). This implies that the BD fraction is much less than 1\% by mass. However, it should be stressed that nobody has yet measured the IMF in the sites which are most likely to be associated with Population III stars.
We have seen that the LMC microlensing results would now seem to exclude a large fraction of BDs in our own halo. Although Honma \& Kan-ya (1998) have presented 100\% BD models, these would require falling rotation curves and most theorists would regard these as rather implausible.

\section{Conclusions}

1) Cosmological nucleosynthesis calculations suggest that there is both baryonic and non-baryonic dark matter. From a theoretical perspective, the latter is better motivated since one cannot be sure that the dark baryons are inside galaxies. The more
conservative conclusion would be that they are contained in an intergalactic medium or in gas within groups or clusters of galaxies.

2) Over the years there have been several observational claims which seem to indicate the existence of MACHOs but these have usually turned out to be false alarms.
Currently the only {\it positive} evidence comes from microlensing observations and the LMC results suggest that white dwarfs may be the best MACHO candidate. 
However, it must be stressed that the mass estimate upon which this inference is based is sensitive to assumptions about the halo model and the large number of arguments which have been voiced against white dwarfs in the past cannot be brushed aside too cavalierly.

3) There are still many viable non-baryonic candidates. As with the baryonic candidates, there have been a number of flase alarms in the past. However, the 
most recent experimental results require that there be at least some hot dark matter in the form of neutrinos and there are also clams to have detected cold dark matter in the form of WIMPs.

4) We have seen that there are many important {\it constraints} on both baryonic and non-baryonic dark matter candidates from a wide variety of other astrophysical effects. Until there is a definite detection, therefore, the best strategy is to proceed by {\it eliminating} candidates, on the Sherlock Holmes principle that whatever candidate remains, however implausible, must be correct.
\vspace{1pc}

\re
1. Alcock, C.  et al. 1997, ApJ 486, 697
\re
2. Alcock, C. et al. 2000, astro-ph/0001272
\re
3. Alexander, T. 1995, MNRAS 274, 909
\re
4. Alfonso, A. et al. 1997, ApJ 99, L12
\re
5. Ashman, K.A. 1990, MNRAS 247, 662
\re
6. Ashman, K.A., Carr, B.J. 1991, MNRAS 249, 13
\re
7. Baganoff, F.K., Malkan, M.A. 1995, ApJ 444, L13
\re
8. Bahcall, J.N., Flynn, C., Gould, A. 1992, ApJ 389, 234
\re
9. Bergstrom, L. 1999, in Cosmo-98, ed. D.Caldwell, in press (astro-ph/9902172)
\re
10. Bond, J.R., Carr, B.J., Hogan, C.J. 1991, ApJ 367, 420
\re
11. Burles, S., Tytler, D. 1998, ApJ 499, 699; 507, 732
\re
12. Burles, S. et al. 1999, Phys. Rev. Lett. 82, 4176
\re
13. Caldwell, D. O. 1999, in The Identification of Dark Matter, ed. N.J.C. Spooner \& V. Kudryavtsev,  p 527 (World Scientific)
\re
14. Carr, B.J. 1994, ARA\&A 32, 531
\re
15. Carr, B.J., Lacey, C.G. 1987, Ap.J 316, 23
\re
16. Carr, B.J., Sakellariadou, M. 1998, ApJ 516, 195
\re
17. Carr, B.J., Bond, J.R., Arnett, W.D. 1984, ApJ 277, 445
\re
18. Carswell R.F. et al. 1994,  MNRAS 268, L1  
\re
19. Cen, R., Ostriker, J.P. 1999, ApJ 514, 1 
\re
20. Charlot, S., Silk,  J. 1995, ApJ 445, 124
\re
21. Chary, R., Wright, E.L., 1998, in 3rd Stromlo Symposium, ed. B.K.Gibson et al.
(ASP Conference Series)  (astro-ph/9811324)
\re
22. Copi, C.J., Schramm, D.N., Turner, M.S. 1995, Phys.Rev.Lett. 75, 3981
\re
23. Covi, L., Kim, J.E., Roszkowski, L. 1999, Phys. Rev. Lett. 82, 4180.
\re
24. Cr\'ez\'e, M. et al. 1998, A \& A 329 920
\re
25. Crawford, M., Schramm, D.N. 1982, Nature 298, 538
\re
26. Davis, R.L., Shellard, E.P.S. 1989, Nuc. Phys. B. 324, 167
\re
27. De Paolis, F. et al. 1995, A \& A 295, 567; 299, 647
\re
28. Ellis, J. 2000, Phys. Scripta T85, 221
\re
29. Ellis, J., Lopez, J., Nanopoulos, D.V. 1990, Phys.Lett. B. 247, 257
\re
30. Elson, R.A. et al. 1999, in Stellar Populations in the Magellanic Clouds, in press 
\re
31. Feltzing, S., Gilmore, G., Wyse, R.F.G. 1999, ApJ 516, L17 
\re
32. Fich, M., Tremaine, S. 1991, ARA\&A 29, 409
\re
33. Field, B.D., Freese, K., Graff, D.S. 2000, ApJ 534, 265 
\re
34. Fixsen, D.J. et al. 1998, ApJ 508, 123
\re
35. Flynn, C., Fuchs, B. 1995, MNRAS 270, 471
\re
36. Fukugita, M., Hogan, C.J., Peebles, P.J.E. 1998, ApJ 503, 518
\re
37. Gawiser, E., Silk, J. 1998, Science 280, 1405
\re
38. Gibson, B.K., Mould, J.R. 1997, ApJ 482, 98
\re
39. Gilmore, G. 1999, in The Identification of Dark Matter, ed. N.J.C. Spooner \& V. Kudryavtsev,  p 121 (World Scientific) 
\re
40. Gilmore, G., Unavane, M. 1998, MNRAS 301, 813
\re
41. Gott, J.R. 1981, ApJ 243, 140
\re
42. Graff, D.S., Freese, K., Walker, T.P., Pinsonneault, M.H. 1999, ApJ 523, L77
\re
43. Green, A. 2000, ApJ 537, 708
\re
44. Hansen, B.M.S. 1998, Nature, 394 860
\re
45. Hawkins, M.R.S. 1993, Nature 366, 242
\re
46. Hawkins, M.R.S. 1996, MNRAS 278, 787
\re
47. Hawkins, M.R.S. 1999, in The Identification of Dark Matter, ed. N.J.C. Spooner \& V. Kudryavtsev,  p 206 (World Scientific)
\re
48. Hodgkin, S.T. et al. 2000, Nature 403, 57 
\re
49. Hogan, C.J. 1999, in Inner Space/Outer Space II; astro-ph/9912107 
\re
50. Honma, M., Kan-ya, Y. 1998, ApJ 503, L139
\re
51. Ibata, R.A., Richer, H.B., Gilliland, R.L., Scott, D. 1999, ApJ 524, L95 
\re
52. Ibata, R.A. et al. 2000, astro-ph/0002138
\re
53. Irwin, M.J. et al. 1989, AJ 98, 1989
\re
54. James, P., Casali, M.M. 1996, Spectrum 9, 14
\re
55. Jedamzik, K. 1997, Phys.Rev.D. 55, R5871 
\re
56. Kerins, E., Carr, B.J. 1994, MNRAS 266, 775
\re
57. Kolb, E.W. 2000, Phys. Scripta T85, 231
\re
58. Kuijken, P., Gilmore, G. 1989, MNRAS 239, 571; 605, 651
\re
59. Kroupa, P., Tout, C., Gilmore, G. 1993, MNRAS 262, 545
\re
60. Lacey, C.G., Ostriker, J.P. 1985, ApJ 299, 633
\re
61. Lequeux, J. et al. 1996, A \& A 312, L1 
\re
62. Lequeux, J. et al. 1998, A \& A 334, L9
\re
63. Lewis, G.F. et al. 1998, MNRAS 295, 573
\re
64. Madau, P., Pozzetti, L. 2000, MNRAS 312, L9 
\re
65. Mathews, G.J., Schramm, D.N., Meyer, B.S. 1993, ApJ 404, 476
\re
66. Paczynski, B. 1986, ApJ 304, 1; 308, L43
\re
67. Paczynski, B. 1996, ARA\&A 34, 419
\re
68. Persic, M., Salucci, P. 1992, MNRAS 258, 14P
\re
69. Preskill, J., Wise, M.B., Wilczek, F. 1983, Phys. Lett. B 120, 127
\re
70. Roszkowski, L. 1999, in Cosmo-98, ed. D.Caldwell, in press (hep-ph/9903467)
\re
71. Rudy, R.P. et al. 1997, Nature 387, 159
\re
72. Rugers, H., Hogan, C.J. 1996, ApJ 459, L1
\re
73. Ryu, D., Olive, K.A., Silk, J. 1990, ApJ 353, 81
\re
74. Sackett, P.D. et al. 1994, Nature 370, 441
\re
75. Schmidt , R., Wambsganss, J. 1998, A \& A 335, 379
\re
76. Shellard, E.P.S. 1990, in Formation and Evolution of Cosmic Strings, 
ed. Gibbons, G.W., Hawking, S.W. \& Vachaspati, V. (Cambridge University Press)
\re
77. Shellard, E.P.S., Battye, R.A. 1998, Phys. Rep. 307, 227.
\re
78. Smecker, T.A., Wyse, R.F.G. 1991, ApJ 372, 448 
\re
79. Songaila, A. et al. 1994, Nature 368, 599  
\re
80. Tamanaha, C.M., Silk, J., Wood, M.A., Winget, D.E. 1990, ApJ 358, 164
\re
81. Thomas, P., Fabian, A.C. 1990, MNRAS 246, 156
\re
82. Turner, M.S. 1999, in Physics in Collision, ed. M.Campbell \& T.M.Wells (World Scientific)
\re
83. Tytler, D., Fan, X.M., Burles, S. 1996, Nature 381, 207
\re
84. Walker, M.A. 1999, MNRAS 306, 504; 308, 551
\re
85. Webb, J.K. et al. 1997, Nature 388, 250
\re
86. Webster, R.L. et al. 1991, AJ 102, 1939 
\re
87. Weinberg, D.H. et al. 1997, ApJ 490, 564 
\re
88. Yost, S.A. et al. 1999, ApJ in press; astro-ph/9908364
\re
89. Zaritsky, D. et al. 1993, ApJ 405, 464
\re
90. Zheng, Z. et al. 1999, AJ 117, 2757
\re
91. Ziaeepour, H. 1999, in The Identification of Dark Matter, ed. N.J.C. Spooner \& V. Kudryavtsev,  p 106 (World Scientific)

%

\chapter*{ Entry Form for the Proceedings }

\section{Title of the Paper}
Baryonic and Non-Baryonic Dark Matter.

{\Large\bf %
\*** 
}

\section{Author(s)}

I am the author of this paper.

\newcounter{author}
\begin{list}%
{Author No. \arabic{author}}{\usecounter{author}}

\item %
\begin{itemize}
\item Full Name:                Bernard John Carr 
\item First Name:               Bernard 
\item Middle Name:              John 
\item Surname:                  Carr 
\item Initialized Name:         B. J. Carr 
\item Affiliation: Queen Mary \& Westfield College, University of London 
\item E-Mail:                   B.J.Carr@qmw.ac.uk 
\item Ship the Proceedings to:  Astronomy Unit, Queen Mary \& Westfield College, Mile End Road, London E1 4NS, England 
\end{itemize}

\end{list}

\end{document}